\documentclass[fleqn,10pt]{wlscirep}
\usepackage[utf8]{inputenc}
\usepackage[T1]{fontenc}
\usepackage{textcomp}
\title{Explainable AI-Driven Neural Activity Analysis in Parkinsonian Rats under Electrical Stimulation}

\author[a,1]{Hanseul Choi}
\author[b,1]{Gaeun Kim}
\author[b,c]{Sunggu Yang}
\author[b,c]{Eunha Baeg}
\author[b,c]{Donggue Kim}
\author[a]{Seongwon Jin}
\author[d,*]{Sangwon Byun}
\author[a,c,*]{Jibum Kim}

\affil[a]{Department of Computer Science and Engineering, Incheon National University, Yeonsu-gu, 22012, Incheon, Republic of Korea}
\affil[b]{Department of Nanobioengineering, Incheon National University, Yeonsu-gu, 22012, Incheon, Republic of Korea}
\affil[c]{Center for Brain-Machine Interface, Incheon National University, Yeonsu-gu, 22012, Incheon, Republic of Korea}
\affil[d]{Department of Electronics Engineering, Incheon National University, Yeonsu-gu, 22012, Incheon, Republic of Korea}
\affil[1]{Hanseul Choi and Gaeun Kim are contributed equally to this work.}
\affil[*]{Corresponding to: Sangwon Byun (swbyun@inu.ac.kr) or Jibum Kim (jibumkim@inu.ac.kr)}

\begin{abstract}
Parkinson's disease (PD) is characterized by motor dysfunction and abnormal neural oscillations. This study combined graphene-based cortical stimulation with an explainable artificial intelligence (XAI) framework to investigate spatial and frequency-related features contributing to pre- and post-stimulation classification in Parkinsonian rats. Electrocorticogram (ECoG) signals were recorded before and after stimulation and classified using EEGNet and DeepConvNet. Layer-wise relevance propagation (LRP) and DeepLIFT were applied to identify frequency bands and recording channels contributing to the classification. The anatomical locations of these channels were mapped onto a cortical map. XAI analyses were performed separately for each animal to examine individual attribution patterns. Our XAI analysis revealed frequency-band- and region-specific contributions to pre- and post-stimulation classification in Parkinsonian rats. Our approach provided complementary insights into the features contributing to pre- and post-stimulation ECoG classification alongside PSD-based analyses. These findings highlight the potential of combining graphene-based cortical stimulation with explainable deep learning to identify classification-relevant features across frequency bands and cortical areas in pre- and post-stimulation recordings.
\end{abstract}
\begin{document}
\flushbottom\maketitle\thispagestyle{empty}


\section*{1. Introduction}

Parkinson's disease (PD), affecting approximately 1\% of individuals over the age of 60, is a neurodegenerative disorder primarily caused by the degeneration of the dopaminergic system \cite{ref1}. PD patients experience a range of motor dysfunctions, such as tremors, rigidity, bradykinesia, and postural instability \cite{ref2}. A popular contemporary clinical treatment for PD is deep brain stimulation (DBS), i.e., electrically stimulating the subthalamic nucleus and globus pallidus interna to alleviate motor symptoms \cite{ref3,ref4}. However, although DBS can reduce symptoms such as tremors, it is an invasive method. DBS can damage non-targeted brain regions during electrode implantation; furthermore, removing deeply implanted electrodes poses considerable challenges. With these concerns in mind, less invasive therapeutic approaches and methods for real-time monitoring of PD are needed \cite{ref32}.

One of the key features of PD is abnormal oscillatory neural activity, particularly in high-frequency bands, such as \(\beta\) and \(\gamma\), which show increased activity \cite{ref9}. Interestingly, electrical stimulation such as DBS has been shown to reverse these patterns by reducing high-frequency oscillations \cite{ref10,ref11}, suggesting that neural activity can be tracked to evaluate the therapeutic potential of electrical stimulation in modulating abnormal brain activity. Therefore, less-invasive approaches that enable continuous monitoring of brainwave changes and corresponding stimulation therapy may contribute substantially to the treatment of PD.

Cortical surface stimulation has also been investigated as a complementary approach for modulating motor circuit activity in PD, with recent evidence demonstrating improvements in gait-related motor deficits in 6-OHDA-induced hemiparkinsonian rats \cite{ref27}. To enable cortical stimulation and monitoring of neural activity, we have developed graphene-based electrodes that are biocompatible, electrically conductive, thermally stable, and mechanically flexible \cite{ref5}. These properties make graphene electrodes well-suited for recording neural activity related to brain disorders and brain mapping, as they minimize tissue damage and inflammatory responses after implantation \cite{ref5,ref6,ref7,ref8}. Furthermore, our graphene electrode arrays can be integrated with wireless microprocessors so that they can measure brain activity and deliver electrical stimulation without movement restrictions. This combined wireless system can offer a less traumatic procedure, allowing easier real-time monitoring and treatment of brain disorders.

To fully harness this potential, it is essential to analyze brain activity recorded through these advanced electrodes accurately. Traditionally, the analysis of brain activity, typically through electroencephalogram (EEG) or electrocorticogram (ECoG) signals, has relied on statistical methods. These approaches have been extensively used in PD research, including studies that reveal abnormal oscillatory activity in high-frequency bands \cite{ref9}. Deep learning allows signals to be classified without requiring handcrafted summary features as model inputs. XAI refers to methods that explain the information contributing to the decisions of a trained model. Previous EEG studies have used XAI to identify channels, brain regions, and signal features that contribute to classification \cite{ref33,ref12,ref34}. In particular, Borra et al. performed XAI-based analyses separately for individual patients to identify classification-relevant EEG features \cite{ref33}. These studies show that XAI can provide information on important spatial and signal features, allowing researchers to identify channels with features that contribute to classification. These studies provide a basis for applying XAI to examine the spectral and spatial information contributing to pre- and post-stimulation ECoG classification in Parkinsonian rats. However, the application of XAI to identify features distinguishing pre- and post-stimulation ECoG recordings in Parkinsonian rat models remains relatively unexplored.

Our study aims to address existing gaps by exploring the feasibility of applying XAI methods to analyze ECoG signals acquired using graphene-based electrodes in a Parkinsonian rat model. To assess whether our framework relies on specific classifiers or XAI methods, we systematically tested various combinations to verify its robustness. Specifically, we employ EEGNet, a lightweight convolutional neural network (CNN) architecture designed for EEG classification tasks \cite{ref16}, and DeepConvNet, an additional CNN-based classifier \cite{ref30}. We apply layer-wise relevance propagation (LRP) and DeepLIFT to identify the frequency bands and recording channels contributing to pre- and post-stimulation classification \cite{ref22,ref31}. LRP propagates relevance scores from the model output to the input layer, whereas DeepLIFT assigns contributions by comparing activations with those obtained from a reference input.

In this study, we apply our XAI-based framework to examine the features contributing to pre- and post-stimulation classification in the Parkinsonian rat model, using graphene-based electrodes for signal acquisition. By interpreting these XAI-based findings alongside results from PSD-based statistical analyses, we aim to provide complementary insights into PD-related neural activity. Our study combines graphene-based cortical stimulation with an explainable deep learning framework to characterize the frequency-band and cortical-area features contributing to pre- and post-stimulation ECoG classification.

\subsection*{2.1. Materials and Methods}

\begin{figure}[!htbp]
\centering
\includegraphics[width=.78\linewidth,trim={0.0000bp 56.2569bp 6.9975bp 10.2477bp},clip]{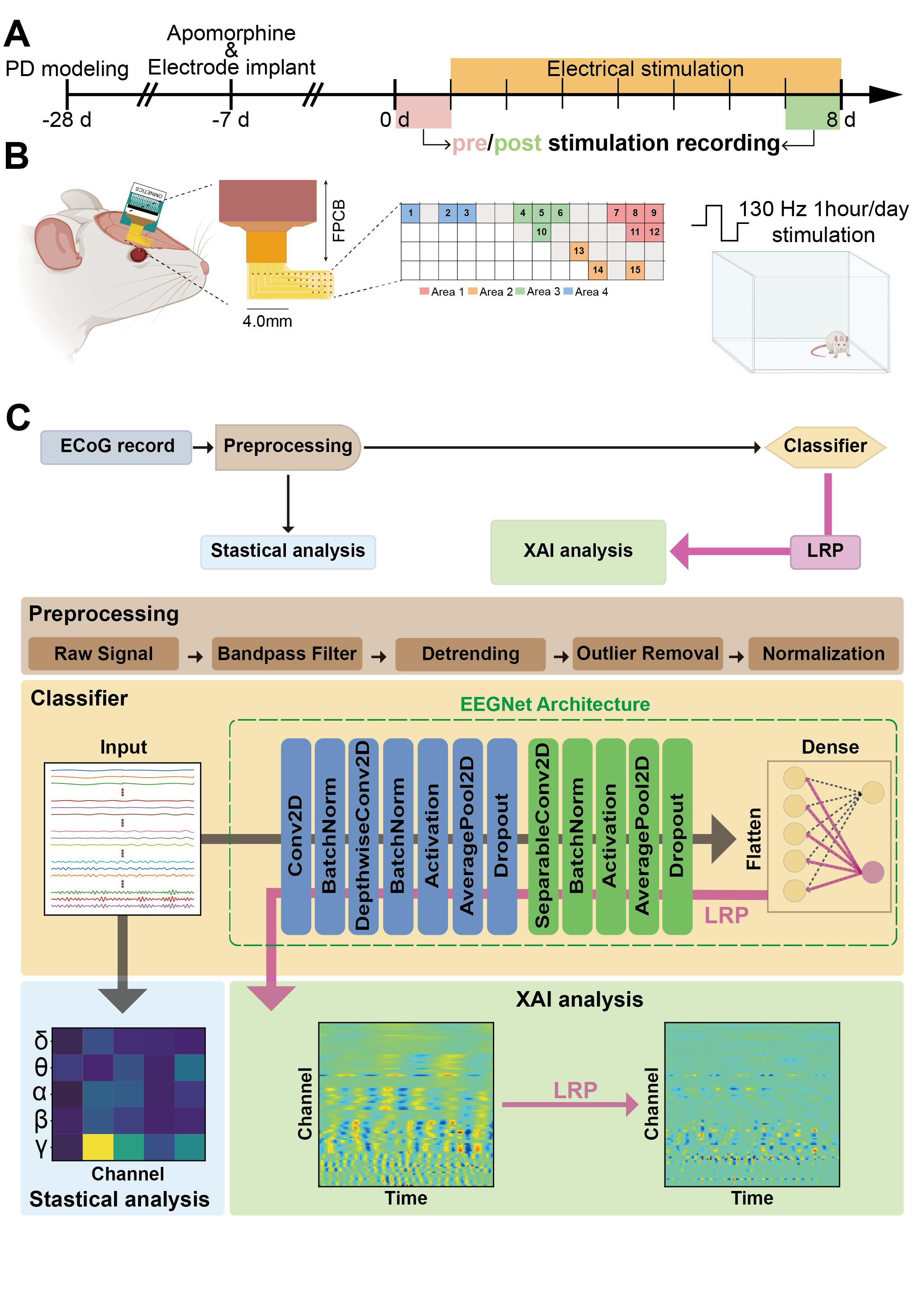}
\caption{Overall experimental procedure and analysis framework. (A) Timeline of PD model induction, apomorphine testing, electrode implantation, electrical stimulation, and pre-/post-stimulation ECoG recordings, with time expressed in days. (B) Cortical electrode placement, electrode grouping into Areas 1--4, stimulation waveform, and recording setup for freely moving rats. (C) Overview of ECoG preprocessing, classification, LRP analysis, and PSD-based statistical analysis, with EEGNet shown as the representative classifier.}
\label{fig:fig1}
\end{figure}

\subsubsection*{2.1.1. Animals}

Each of four adult male Sprague Dawley rats (260--300 g) was labeled with an identifier: PD1, PD2, PD3, and PD4. The rats were accommodated in a facility whose temperature was set at 25\(^{\circ}\mathrm{C}\) and subjected to a 12-hour light/dark cycle. All animal procedures were approved by the Institutional Animal Care and Use Committee at Incheon National University (Approval Code: INU-ANIM-2017-08) in accordance with their guidelines, and all experiments were conducted in compliance with the ARRIVE guidelines.

\subsubsection*{2.1.2. 6-OHDA-induced hemiparkinsonian rat model}

To establish a pharmacologically induced PD model in animals, 6-hydroxydopamine (6-OHDA) was stereotactically injected into the dopaminergic neurons of the medial forebrain bundle. Under isoflurane anesthesia, a single burr hole was made in rats at coordinates \(-\)3.84 mm anterior and \(-\)1.4 mm lateral relative to the bregma. A solution containing 2.94 mg of 6-OHDA in saline and 0.1\% ascorbic acid was injected using a Hamilton syringe (26G needle) with a micro-pump (LMS Korea, Seongnam, Gyeonggi, Republic of Korea) at a depth of 8.5 mm from the dura. The infusion was administered at a rate of 0.5 \(\mu\)l/min over a period of 8 minutes into one hemisphere, allowing for a straightforward comparison with the non-affected hemisphere, which served as the control. Three weeks following 6-OHDA administration, the rats were injected with apomorphine (0.5 mg/kg), and an apomorphine-induced rotation test was performed for 30 minutes to confirm the induction of PD. Rats that exhibited at least three full-body rotations per minute were considered to have successfully developed the model.

\subsubsection*{2.1.3. Graphene-based electrode fabrication}

We utilized the same technique employed in our previous studies to create a graphene-based electrode array \cite{ref4,ref12,ref7,ref8,ref5}. Monolayer graphene grown on Cu foil through chemical vapor deposition (CVD) was stacked into four layers following the chemical doping process. A 1.2-\(\mu\)m-thick polyimide (PI) layer was deposited by spin-coating onto a copper carrier substrate, forming a flexible printed circuit board (FPCB). For the interconnects, a layer of Cr/Au (3 nm/40 nm) was applied onto the PI layer using thermal evaporation followed by photolithography. The four-layer graphene was then transferred onto the PI substrate, aligned with the Au contact tracks, and shaped using photolithography combined with oxygen plasma treatment. SU-8 was used to insulate the undesired areas of the Au tracking metal. The graphene was subsequently patterned and etched using photolithography and oxygen plasma, followed by nitric acid chemical doping.

\subsubsection*{2.1.4. Experimental designs for electrotherapy and recording}

In animals confirmed to have PD via the apomorphine-induced rotation test, a graphene-based electrode array integrated with an Omnetics connector (Gbrain, Incheon, Korea) was surgically implanted onto the motor cortex of the affected hemisphere. The array was encapsulated by the skull and securely anchored using dental cement. Electrical stimulation was delivered directly to the motor cortex, with stimulation sessions beginning one-week post-implantation. The ECoG recordings analyzed in this study were obtained from the animals used in our previous study \cite{ref27}, which evaluated behavioral outcomes following cortical stimulation. The 1 h/day stimulation protocol for 8 consecutive days was selected based on the experimental paradigm reported by Shin et al. \cite{ref27}, in which differences in behavioral and neurophysiological responses between cortical stimulation and DBS became evident within the first week of treatment. The Parkinsonian rats received an hour of electrical stimulation (biphasic, 130 Hz, 66.6 \(\mu\)s, 300 \(\mu\)A) at the same time each day during the daytime for 8 consecutive days (Figure 1A) to examine ECoG changes following repeated cortical stimulation. Stimulation was controlled using an RHS stim/recording controller (INTAN Technologies, Los Angeles, CA, USA), which was connected to the implanted electrode via a head stage and an 8-Mbit/s serial peripheral interface cable (INTAN Technologies, Los Angeles, CA, USA). Figure 1B shows a rat implanted with the head stage and electrodes.

Neural activity was recorded using a wireless device, as described below. ECoG signals from the motor cortex were recorded for 30 minutes prior to the first stimulation as well as after the final stimulation using a 15-channel graphene-based electrode array. The channels were grouped into four putative motor cortical areas based on the predefined motor cortex map (Figure 1B): Area 1 (channels 7, 8, 9, 11, and 12) corresponding to the face, Area 2 (channels 13, 14, and 15) corresponding to the forelimb, Area 3 (channels 4, 5, 6, and 10) corresponding to the trunk, and Area 4 (channels 1, 2, and 3) corresponding to the hindlimb \cite{ref26}. All recordings were conducted in freely moving, awake Parkinsonian rats within a 43\(\times\)33\(\times\)30 cm cage placed inside a Faraday cage, without food, water, or objects present during measurements. (Figure 1B). Electrode implantation was performed using standardized stereotaxic coordinates and the same electrode configuration across all animals (Supplementary Figure S1). Motor threshold testing was performed after implantation, and stimulation-induced motor responses were confirmed, providing functional confirmation that the electrode array was targeting the motor cortical region.

\subsubsection*{2.1.5. Wireless device}

The proposed neuroprosthetic device delivers short-latency electrical stimulation driven by physiological signals and operates autonomously via wireless telemetry. Designed specifically for animal models, the device is optimized for minimal size, weight, and power consumption. The complete wireless headstage weighs 3.8 g, including the electronic components and battery, and integrates critical functions such as stimulation, recording, telemetry, power management, and on-device computing through a dedicated microprocessor. The device consumes approximately 39 mW, corresponding to a current draw of 10.5 mA at 3.7 V. It is powered by a 150 mAh rechargeable battery, providing approximately 10 hours of operation. It utilizes Bluetooth Low Energy (BLE) in the 2.4 GHz band for wireless data transmission, ensuring high data integrity through sample sequence values that maintain precise timing, even during intermittent connection losses.

\subsection*{2.2. Signal Preprocessing}

Given that brain wave data is susceptible to various noise sources, such as environmental noise or the subject's movement, it is necessary to remove noise before analysis. Given that the rats were freely moving during the measurements, the likelihood of movement-induced noise was higher compared to stationary or anesthetized subjects. Therefore, in addition to the commonly used bandpass or notch filters, we incorporated two more preprocessing steps: detrending and outlier removal \cite{ref19}.

The 30-min long wirelessly recorded signals were sampled at 512 Hz. In this study, we divided the signals into five commonly investigated frequency bands: \(\delta\) (0.4--4 Hz), \(\theta\) (4--8 Hz), \(\alpha\) (8--13 Hz), \(\beta\) (13--30 Hz), and \(\gamma\) (30--50 Hz). To evaluate the contribution of each frequency band to classification performance, the signals from 15 channels were replicated five times, resulting in multi-band datasets. These replicated data from each channel were subjected to bandpass filters with five different ranges: \(\delta\), \(\theta\), \(\alpha\), \(\beta\), and \(\gamma\). Notably, the \(\delta\) band typically includes frequencies between 0 Hz and 4 Hz, but in our study, the filter range was adjusted to 0.4--4 Hz to reduce movement-induced noise. After filtering, linear-type detrending was applied. Outliers were identified using an absolute-amplitude z-score threshold of 3, with the reference mean and standard deviation estimated from training data only, and replaced by cubic interpolation within each segment. The signals were standardized using the mean and standard deviation estimated from training data only. The same parameters were applied to the validation and test data. Preprocessing was performed using Python 3.11.16 and SciPy 1.13.1 on an AMD Ryzen 7 9700X 8-core processor under Linux/WSL2, with approximately 15.2 GB of RAM visible to the WSL environment.

\subsection*{2.3. Deep learning}

Model training and XAI computations were performed using Python 3.11.16 and TensorFlow 2.21.0 with tf-keras 2.21.0 under Linux/WSL2. The computer was equipped with an AMD Ryzen 7 9700X 8-core processor and an NVIDIA GeForce RTX 5060 GPU with 8 GB of memory.

\subsubsection*{2.3.1 EEGNet}

EEGNet is a compact and efficient CNN-based classifier designed for EEG signal analysis. It achieves superior performance compared to traditional methods while noticeably reducing the number of trainable parameters, making it ideal for analyzing EEG data effectively. EEGNet architecture has been extensively validated across various studies, demonstrating its efficacy in diverse scenarios owing to its lightweight design and ease of implementation \cite{ref20,ref21}.

In this study, the EEGNet classifier processed two-dimensional (2D) ECoG signals as inputs, comprising temporal (horizontal axis) and spatial (vertical axis) features, and classified the data into two states: pre-stimulation and post-stimulation. Each sample of data, six-second long, was segmented without overlap from a preprocessed 30-min long signal. Each sample included six-second long segments from multi-band datasets, which were composed of five frequency bands across 15 channels. To reduce the influence of temporal correlations between nearby samples assigned to different data splits, we employed five-fold contiguous-block cross-validation within each animal, partitioning the recordings according to temporal order using an approach similar to those described in previous EEG studies \cite{ref28,ref29}. Each pre- and post-stimulation recording was divided into five contiguous blocks according to temporal order. In each fold, one block from each recording was used for testing, one for validation, and the remaining three for training. Each block was used once for testing across the five folds. To ensure a minimum temporal gap of 30 seconds between different training, validation, and test data sets, epochs overlapping with intervals of 15 seconds on both sides of each data-set boundary were excluded. As described in Section 2.2, outlier reference statistics and standardization parameters were estimated from training data only and applied to the validation and test data. For model training, we used an Adam optimizer with a fixed learning rate of 0.001. A batch size of 16 was used. We trained for 100 epochs using training data. The model with the smallest validation loss was selected.

One of the best strategies to build a classifier is to gather a large number of subjects and train the classifier in a subject-independent manner. This approach allows the classifier to learn general features by using all available data. Unfortunately, it was challenging to increase the number of subjects in this study because of difficulties in modeling PD in rats and the high cost of electrodes. Consequently, we trained a classifier for each individual subject using a subject-dependent approach and analyzed whether each classifier showed consistent trends. The average accuracy of the 5-fold test for each of the four rats was over 98\%, indicating exceptionally high performance.

The EEGNet architecture adopted in this study is depicted in Figure 1C. This model comprises three primary blocks. The initial layer of the first block is a 2D convolutional layer, which generates feature maps capturing characteristics from various frequency bands through convolution properties. Here, the model acquires knowledge of features across diverse frequency ranges, followed by batch normalization. In the subsequent depthwise convolution layer, spatial filters are learned for each feature map individually, enabling the model to learn frequency-specific spatial filters. This is succeeded by another batch normalization, an activation layer, average pooling layer, and dropout. The second block comprises a separable convolution layer, subsequently followed by batch normalization, activation, average pooling, and dropout. The separable convolution layer combines depthwise convolution and pointwise convolution, with the former summarizing the features for each feature map and the latter combining these summarized features. The final block, the classifier block, involves flattening the output from the second block and passing it through a dense network to classify the data \cite{ref16}.

\subsubsection*{2.3.2 DeepConvNet}

DeepConvNet was included as an additional CNN-based classifier to assess whether the XAI results depended on the choice of network architecture. We used a reduced two-block variant adapted from the DeepConvNet architecture proposed by Schirrmeister et al. \cite{ref30}.

The model classified pre- and post-stimulation ECoG recordings using the same inputs and preprocessing as EEGNet. Each input contained a non-overlapping 6-second segment with five frequency bands across 15 electrodes. We used the same five-fold contiguous-block cross-validation procedure described in Section 2.3.1. Training was performed using the Adam optimizer with a learning rate of 0.001, a batch size of 16, and 100 epochs. The model with the lowest validation loss was selected.

A separate DeepConvNet classifier was trained for each animal using the same training, validation, and test samples as EEGNet. This allowed the two architectures to be compared under the same data partitions. Classification accuracy is reported in Table 1.

The reduced architecture consisted of two convolutional blocks followed by a classification layer. The first block used a temporal convolution with 16 filters and a kernel size of 1 \(\times\) 5, followed by a spatial convolution with 16 filters spanning all 75 electrode--band inputs. Batch normalization, ELU activation, temporal max pooling with a pool size of 1 \(\times\) 2, and dropout at a rate of 0.5 were then applied. The second block used a temporal convolution with 32 filters and a kernel size of 1 \(\times\) 5, followed by the same normalization, activation, max pooling, and dropout operations. An additional temporal average pooling layer with a pool size and stride of 1 \(\times\) 4 was applied before flattening. A dense layer with two outputs and a softmax activation produced the probabilities for the pre- and post-stimulation classes.

\subsubsection*{2.3.3 Layer-wise Relevance Propagation}

We used Layer-wise Relevance Propagation (LRP) \cite{ref22} to determine which channels the models focus on when classifying pre- and post-stimulation in Parkinsonian rats. LRP, a well-known framework in the field of XAI, decomposes the relevance score from the network's output backward through to the input layer. This process analyzes how each input feature contributes to the networks' predictions. When redistributing the relevance score in LRP, the following rule is applied, known as the naive rule \cite{ref23}. The relevance score \({R}_{i}^{l}\) for neuron \(i\) in layer \(l\) is given by the following equation:

\[
{R}_{i}^{l}=\sum_{j} \frac{{z}_{ij}}{\sum_{{i}^{'}} {z}_{{i}^{'}j}  }{R}_{j}^{(l+1)}    \quad\text{with}\quad   {z}_{ij}={x}_{i}^{\left(l\right)}{w}_{ij}^{(l,l+1)}
\]

Here, \(l\) indexes the layer, \(i\) indexes the neurons in layer \(l\), and \(j\) indexes the neurons in layer \(l+1\). The term \({R}_{i}^{l}\) represents the relevance score of neuron \(i\) in layer \(l\). The term \({z}_{ij}\) is the product of the activation value \({x}_{i}^{\left(l\right)}\) of neuron \(i\) in layer \(l\) and the weight \({w}_{ij}^{(l,l+1)}\) from neuron \(i\) in layer \(l\) to neuron \(j\) in layer \(l+1\). This product \({z}_{ij} \)indicates how much neuron \(i\) contributes to the activation of neuron \(j\) in the next layer. The denominator \(\sum_{{i}^{'}} {z}_{{i}^{'}j}\) represents the sum of contributions \(z\) from all neurons \({i}^{'} \)in layer \(l\) to neuron \(j\) in layer \(l+1\), serving as a normalization factor. This normalization ensures that the relevance score is distributed proportionally to the contribution of each neuron. When this redistribution algorithm is applied, it produces a relevance map (or heatmap) that satisfies the conservation property \(\sum_{p} {R}_{p}^{(1)}=  f(x)\), where \({R}_{p}^{\left(1\right)}\) denotes the relevance score for the \(p\)th input feature in the first layer, representing how much the \(p\)th input feature contributes to the overall output of the network. In other words, the total relevance score assigned to all input features is equal to the network's output value.

The \(\epsilon \)-rule is a variant of the naive redistribution rule described above and is expressed as:

\[
{R}_{i}^{l}=\sum_{j} \frac{{z}_{ij}}{\sum_{{i}^{'}} {z}_{{i}^{'}j}+\epsilon \operatorname{sign}(\sum_{{i}^{'}} {z}_{{i}^{'}j} )  }{R}_{j}^{(l+1)}
\]

Here, \(\epsilon >0\) is introduced to prevent the denominator from becoming zero, thus ensuring numerical stability. It also has the effect of reducing the values of neurons with very small contributions, thereby removing noise from the explanation. However, unlike the naive rule, the \(\epsilon \)-rule does not satisfy the conservation property. We used \(\varepsilon\) = 10\(^{-4}\), the default value in the DeepExplain implementation. Varying \(\varepsilon\) from 10\(^{-6}\) to 10\(^{-2}\) resulted in a maximum absolute difference of 0.191\% of total relevance in the normalized relevance proportions for EEGNet, indicating limited sensitivity within the tested range.

When analyzing the relevance map, positive values indicate features that support the model's prediction for the target class, while negative values indicate features that oppose it. Using the absolute values of the relevance scores allows us to emphasize the magnitude of the feature's influence, regardless of whether it is positive or negative. Features with values close to zero have little impact on the model's decision, whereas features with larger absolute values have a higher impact on the model's decision. Therefore, the relevance map with absolute values can indicate how much the model relies on specific features. We divided each channel into five frequency bands through preprocessing and analyzed the importance by channel as well as band. For example, the XAI analysis block of Figure 1C depicts the heat map of input data and the relevance map obtained through LRP. We analyzed relevance maps of correctly classified post-stimulation samples from the held-out test sets to determine the trend of feature importance across frequency bands and channel positions.

\subsubsection*{2.3.4 DeepLIFT}

DeepLIFT was included as an additional XAI method to assess whether the attribution patterns depended on the explanation method. DeepLIFT assigns contributions to input features by comparing their activations with those obtained from a reference input \cite{ref31}.

We applied the DeepLIFT Rescale rule implemented in DeepExplain to EEGNet and DeepConvNet. Attribution was computed for the target-class output before softmax, using a zero reference input in the standardized input space. This reference corresponds to the training-data mean for each standardized input feature.

We analyzed the same correctly classified post-stimulation samples from the held-out test sets used for LRP. Absolute attribution values were summed across time and samples and pooled across the five cross-validation folds for each animal and model. The same aggregation and normalization procedures used for LRP were applied to generate electrode maps and calculate band- and area-level attribution proportions. The DeepLIFT results are presented in Supplementary Figures S3A--D.

\subsection*{2.4. Statistical analysis}

Statistical analyses were performed in Python using SciPy, with animals treated as the independent statistical units (n = 4). Group-level results are presented as mean \(\pm\) SEM across animals. Classification accuracy is presented as mean \(\pm\) SD in Table 1 and as mean accuracy in Table 2, calculated across five held-out test folds within each animal. Pre-/post-stimulation comparisons were evaluated using two-sided paired t-tests. Comparisons involving three or more repeated conditions were evaluated using one-way repeated-measures ANOVA with Greenhouse--Geisser correction, followed by two-sided paired comparisons. Bonferroni correction was applied within each comparison family, as specified in the figure captions. Statistical significance was defined as p < 0.05. Effect sizes were reported as Cohen's dz for paired comparisons and partial \(\eta\)\(^{2}\) for repeated-measures ANOVA. Detailed statistical results, including p-values and effect sizes, are provided in Supplementary Tables S1--S6. Graphs were created using GraphPad\textregistered{} Prism 7 (GraphPad Software Inc., La Jolla, CA). Illustrations were produced using Illustrator CC 2021 (Adobe Inc., San Jose, CA) and Biorender.com. Power spectral density (PSD) estimation was conducted using Welch's method in Python 3.7.12 with SciPy 1.7.3. The band power ratio for each frequency band was calculated by summing the PSD values corresponding to each band and dividing by the total sum of the PSD values.

\section*{3. Results}

Section 3.1 presents the experimental settings, including the classifiers and the input data format. Section 3.2 examines pre- and post-stimulation differences in relative band power using PSD-based statistical analyses. Section 3.3 uses XAI to identify the recording channels, cortical areas, and frequency bands contributing to pre- and post-stimulation classification. Finally, Section 3.4 examines whether the frequency-band relevance patterns vary with the input window length.

\subsection*{3.1 Experimental settings}

To analyze the importance of each frequency band, the ECoG signals from the 15 original channels were each replicated five times, resulting in a total of 75 channel--band combinations. These expanded signals were then subjected to the bandpass filters corresponding to the five main brainwave bands: \(\delta\), \(\theta\), \(\alpha\), \(\beta\), and \(\gamma\). The channels were grouped into four putative motor cortical areas based on the predefined motor cortex map: Area 1 (face), Area 2 (forelimb), Area 3 (trunk), and Area 4 (hindlimb), as described in the Methods section \cite{ref26}.

\begin{table}[!htbp]
\centering\small
\caption{Classification accuracy of EEGNet and DeepConvNet in individual rats. Values show the mean \(\pm\) SD accuracy (\%) across five held-out test folds for each rat using 6-second inputs. DeepConvNet denotes the reduced two-block model.}
\label{tab:table1}
\begin{tabular}{lll}
\toprule[1pt]
Animal & EEGNet (\%) & DeepConvNet (\%) \\
\noalign{\vskip\aboverulesep\begingroup\color[HTML]{808080}\hrule height 0.5pt\endgroup\vskip\belowrulesep}
PD1 & 98.390 \(\pm\) 3.590 & 95.360 \(\pm\) 10.380 \\
PD2 & 100.000 \(\pm\) 0.000 & 99.220 \(\pm\) 1.750 \\
PD3 & 100.000 \(\pm\) 0.000 & 100.000 \(\pm\) 0.000 \\
PD4 & 100.000 \(\pm\) 0.000 & 100.000 \(\pm\) 0.000 \\
\bottomrule[1pt]\end{tabular}
\end{table}

Each input sample was 6 seconds long and segmented without overlap from the preprocessed signals. We employed EEGNet \cite{ref20,ref21} and a reduced two-block DeepConvNet \cite{ref30} to analyze these input data. The EEGNet network structure followed the specifications presented in the original paper \cite{ref16}, and the kernel size for the temporal convolutional layer was set to (1 \(\times\) 256). Four Parkinsonian rats were studied. Given the limited number of subjects, the classifiers were trained subject-dependently. To reduce the influence of temporal correlations between nearby samples assigned to different data splits, five-fold contiguous-block cross-validation within each animal was employed, training five classifiers for each model architecture and rat, as described in Section 2.3. Classification accuracy is reported as the mean \(\pm\) SD across the five held-out test folds in Table 1.

\subsection*{3.2. Statistical analysis based on PSD}

\begin{figure}[!htbp]
\centering
\includegraphics[width=.74\linewidth,trim={1.2506bp 60.2553bp 0.7504bp 8.2530bp},clip]{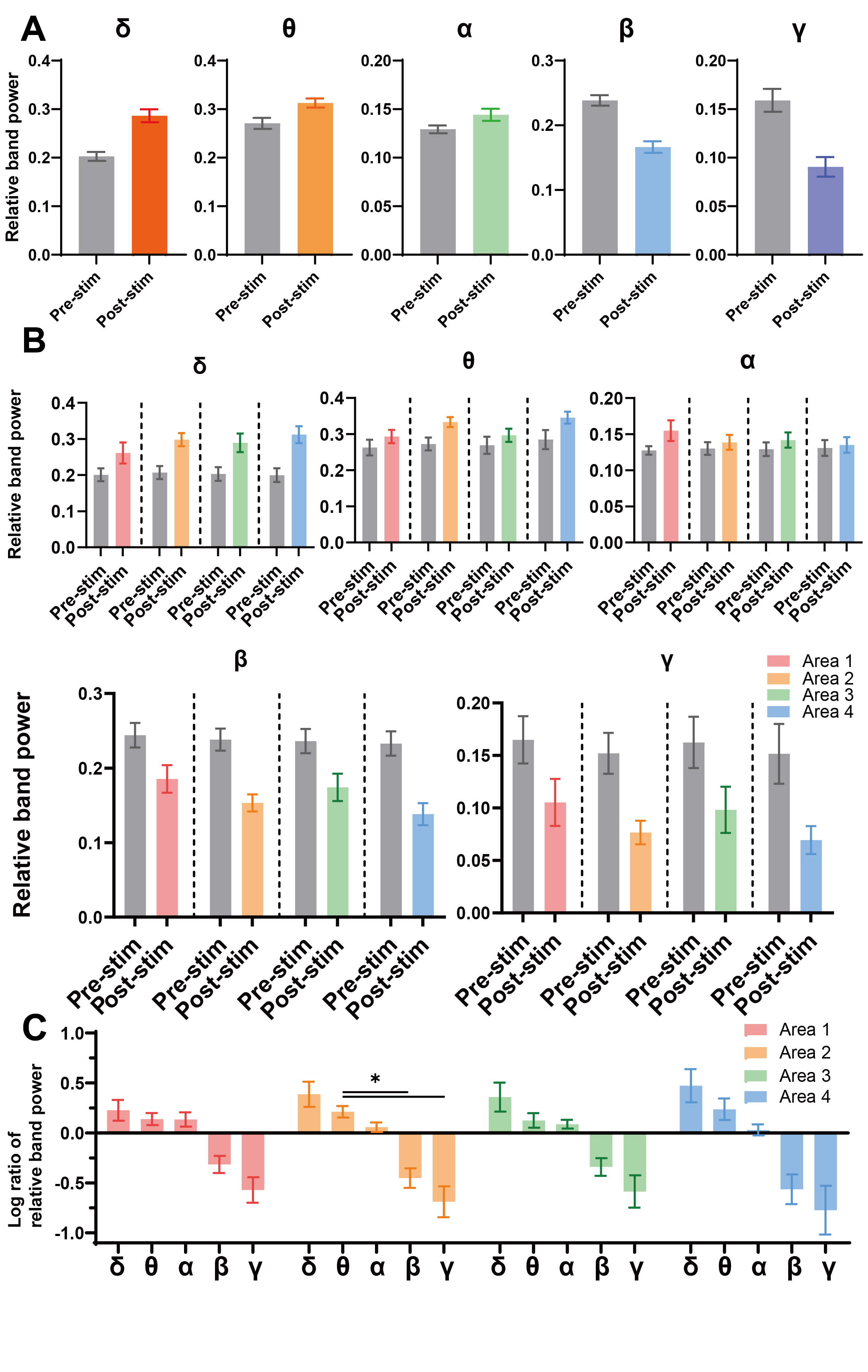}
\caption{Relative band power before and after electrical stimulation. (A) Relative band power averaged across all 15 electrodes within each rat. (B) Relative band power averaged within Areas 1--4, containing 5, 3, 4, and 3 electrodes, respectively. Gray and colored bars indicate pre- and post-stimulation values, respectively. (C) Channel-level ln(post/pre) relative band power averaged within each area; positive and negative values indicate increases and decreases after stimulation. Bars show the mean \(\pm\) SEM across four rats. Pre-/post-stimulation differences in A and B were evaluated using two-sided paired t-tests with Bonferroni correction across the five bands separately for the whole-channel summary and each area. Band effects within each area in C were evaluated using repeated-measures ANOVA with Greenhouse--Geisser correction and Bonferroni-adjusted paired comparisons. *p < 0.05.}
\label{fig:fig2}
\end{figure}

We conducted PSD-based analyses between pre-stimulation (pre-stim) and post-stimulation (post-stim) phases to examine how brainwave activity in different frequency bands changes before and after stimulation in Parkinsonian rats. Animals were treated as the independent statistical units (n = 4). Pre-/post-stimulation differences in Figures 2A and 2B were evaluated using two-sided paired t-tests with Bonferroni correction across the five bands separately for the whole-channel summary and each area. Band effects within each area in Figure 2C were evaluated using repeated-measures ANOVA with Greenhouse--Geisser correction and Bonferroni-adjusted paired comparisons. All PSD data were estimated from preprocessed ECoG data using Welch's method, as described in Section 2.4.

First, we compared pre- and post-stimulation power ratios averaged across all 15 recording channels within each rat. Figure 2A shows the mean \(\pm\) SEM across four rats. No significant differences were observed in mean power ratios between pre- and post-stimulation in any frequency band after Bonferroni correction. The mean power ratio increased in the \(\delta\) band and decreased in the \(\beta\) and \(\gamma\) bands. These decreases were consistent in direction with previous studies that showed decreased neural activity in high-frequency bands following electrical stimulation in Parkinsonian rats \cite{ref9,ref11,ref12}.

Then, a PSD comparison was performed considering the electrode positioning to examine pre- and post-stimulation changes within each cortical area (Figure 2B). We compared power ratios averaged within Areas 1--4, containing 5, 3, 4, and 3 recording channels, respectively, for each frequency band within each rat. Figure 2B shows the mean \(\pm\) SEM across four rats. The trend of changes in power ratios before and after stimulation was consistent with that shown in Figure 2A. After the stimulation, the mean power ratio in the \(\delta\) band increased across all areas compared to pre-stimulation, while those in the \(\beta\) and \(\gamma\) bands decreased across all areas. However, no significant differences were observed in mean power ratios between pre- and post-stimulation in any area or frequency band after Bonferroni correction across the five bands separately within each area.

Finally, for each area, we investigated whether the pre- and post-stimulation differences in power ratios differed among frequency bands (Figure 2C). Band-specific changes were assessed by dividing the post-stimulation power ratio by the pre-stimulation power ratio and taking the natural logarithm. These changes were calculated for each recording channel and averaged within each area for each rat. The resulting values were analyzed using one-way repeated-measures ANOVA with Greenhouse--Geisser correction and Bonferroni-adjusted paired comparisons. The \(\delta\) band exhibited a positive mean value (increase) in all areas, while the \(\beta\) and \(\gamma\) bands showed a negative mean value (decrease) in all areas, similar to the results shown in Figure 2B. Area 2 showed significant differences between the \(\theta\)--\(\beta\) and \(\theta\)--\(\gamma\) bands. No significant overall band effects were observed in the other areas.

In summary, after the stimulation, there was a tendency for decreased power ratios in high-frequency bands, \(\beta\) and \(\gamma\), and increased power ratios in the \(\delta\) band across all areas. However, no significant differences were observed in mean power ratios between pre- and post-stimulation when averaged across all 15 recording channels or separately within each area (Figures 2A and 2B). In contrast, Area 2 showed significant differences between the \(\theta\)--\(\beta\) and \(\theta\)--\(\gamma\) bands in the log-transformed changes in power ratios (Figure 2C).

\subsection*{3.3. XAI-based analysis}

We performed XAI-based analyses by applying LRP and DeepLIFT to EEGNet and DeepConvNet classifiers trained on pre- and post-stimulation ECoG recordings. This approach provided information on the spectral and spatial features contributing to classification and allowed us to examine the information used by the classifiers without requiring handcrafted summary features as model inputs. XAI-based analyses were performed independently for each animal, and the findings were interpreted alongside conventional PSD analysis. Supplementary Figure S2 presents individual input examples illustrating how LRP identifies classification-relevant features. The LRP results for each model are presented in Figures 3--6, and the DeepLIFT results obtained using the same models are presented in Supplementary Figures S3A--D.

For the analyses in Figures 3--6, we analyzed relevance maps of correctly classified post-stimulation samples from the held-out test sets to determine the trend of feature importance across frequency bands and channel positions. These samples are referred to as true positives (TPs). Absolute relevance was summed across time and samples, with the five test folds aggregated within each rat and model. Focusing only on the TPs, this analysis helps identify which features played a key role in the classifier's decision to label a particular sample as post-stimulation.

\begin{figure}[!htbp]
\centering
\includegraphics[width=.80\linewidth,trim={8.7483bp 48.9985bp 17.7528bp 4.2523bp},clip]{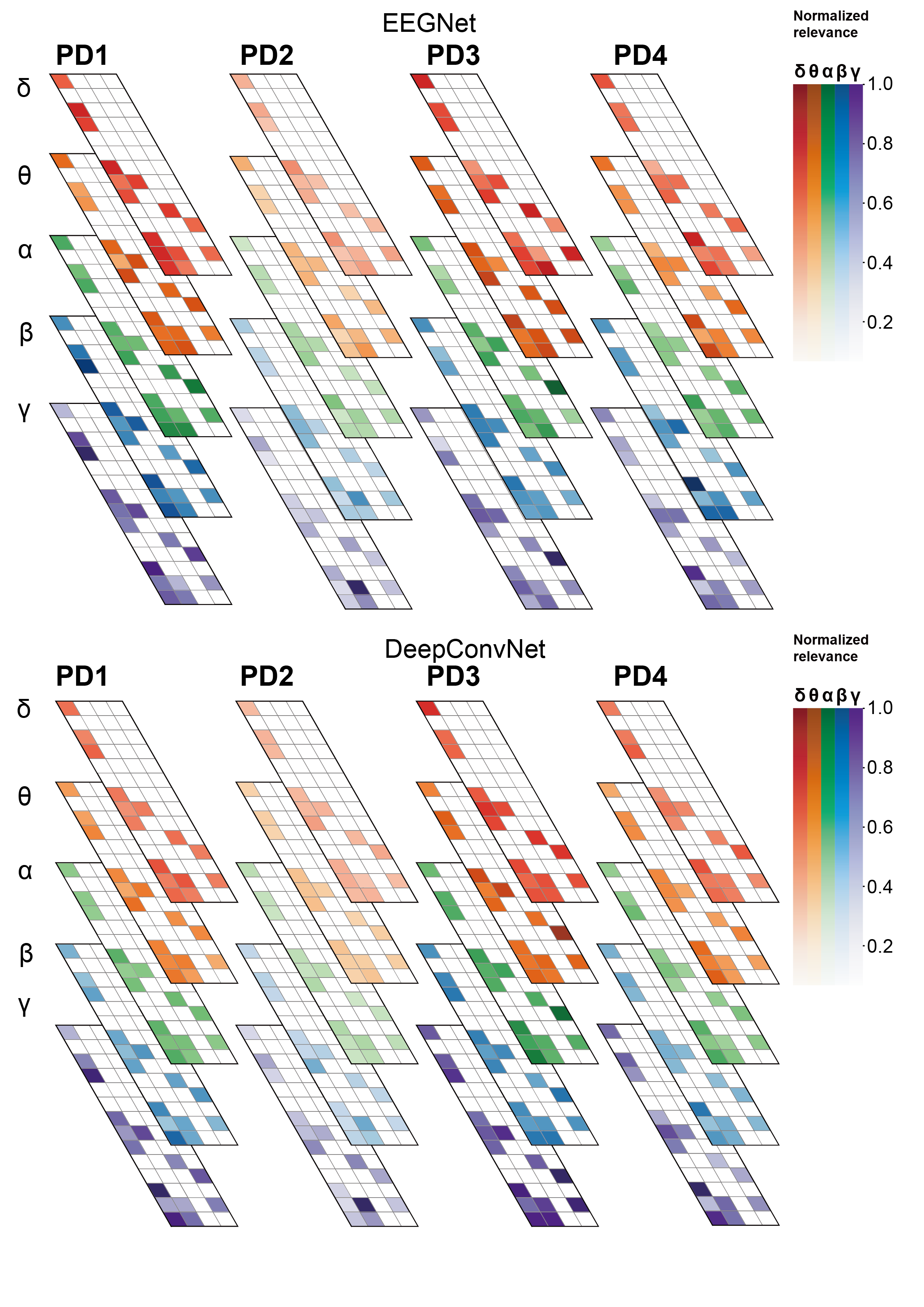}
\caption{Spatial distribution of LRP relevance across electrodes and frequency bands. Results are shown for EEGNet (top) and the reduced two-block DeepConvNet (bottom), with columns corresponding to PD1--PD4. Each grid represents the electrode layout for one frequency band. Each color corresponds to the normalized absolute relevance score in each frequency band. Absolute relevance score from correctly classified post-stimulation test samples was summed across time and samples, then normalized by the maximum of the 75 band \(\times\) electrode values within each rat and model.}
\label{fig:fig3}
\end{figure}

Figure 3 shows the normalized absolute LRP relevance scores across recording channels and frequency bands for EEGNet and DeepConvNet. For each rat and model, the aggregated relevance scores were normalized by the maximum score across all input features. Each grid represents the electrode layout for one frequency band, allowing us to identify channels with relatively high contributions to classification.

The two models assigned relatively high relevance to overlapping channels, particularly in the \(\gamma\) band. Within this band, channels 3 and 7 were the two highest-ranked channels in both models for PD1, although their order differed. Both models assigned the highest relevance to channel 11 in PD2, channel 14 in PD3, and channel 7 in PD4. These shared channels indicate that some spatial features contributing to classification were identified by both network architectures. However, the relevance patterns were not identical across models or rats, suggesting that the identified features reflected both shared and model-dependent characteristics.

\begin{figure}[!htbp]
\centering
\includegraphics[width=.95\linewidth,trim={7.7479bp 367.7708bp 2.0010bp 10.9971bp},clip]{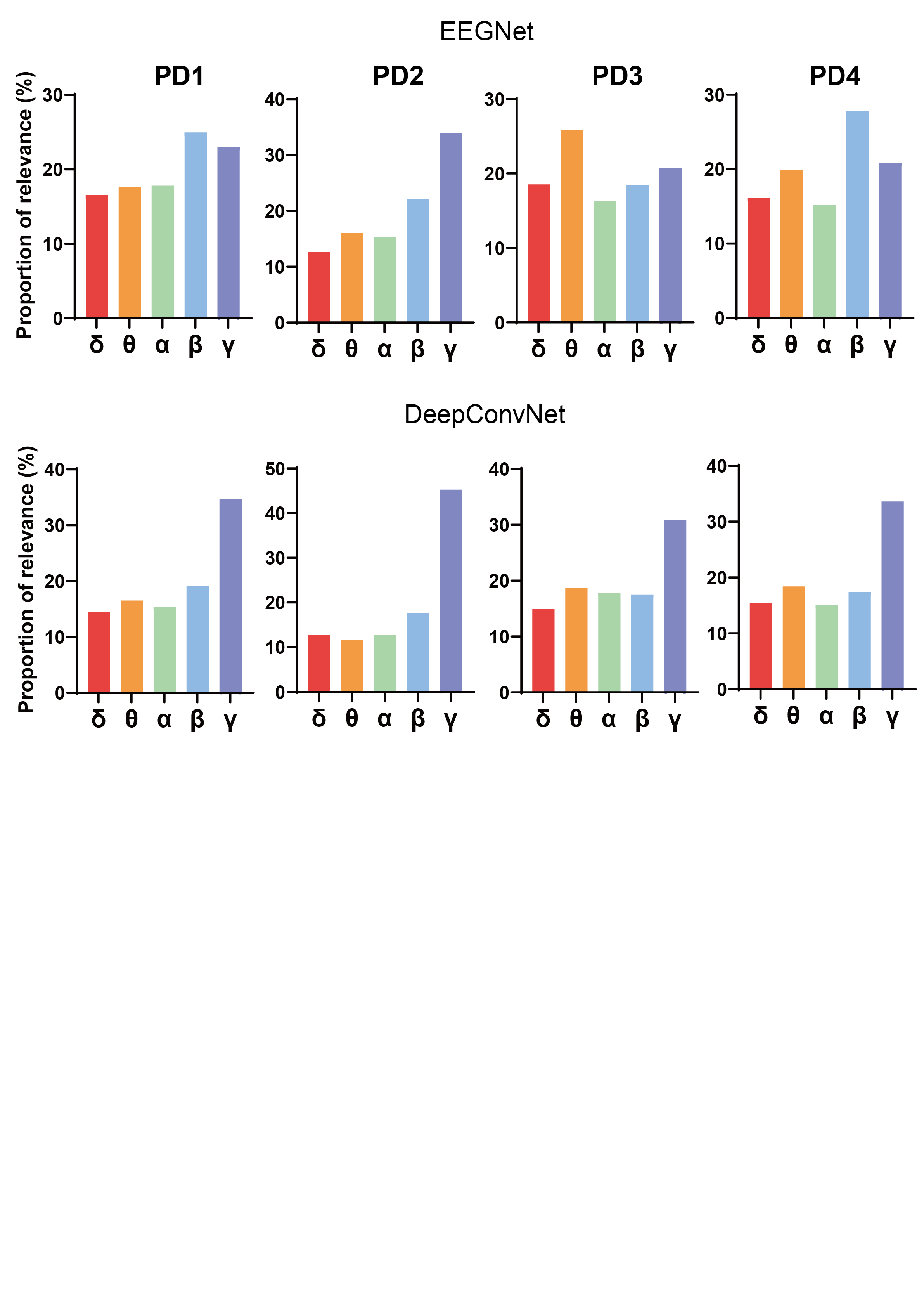}
\caption{Relative LRP relevance score of each frequency band in individual rats. Results are shown for EEGNet (top) and the reduced two-block DeepConvNet (bottom), with columns corresponding to PD1--PD4. Each bar represents the percentage of total absolute relevance assigned to one frequency band in correctly classified post-stimulation test samples from an individual rat. The five band proportions sum to 100\% within each rat and model.}
\label{fig:fig4}
\end{figure}

Figure 4 shows the percentage of total absolute relevance assigned to each frequency band in correctly classified post-stimulation test samples from individual rats. For EEGNet, the highest proportion was observed in the \(\beta\) band in PD1, the \(\gamma\) band in PD2, the \(\theta\) band in PD3, and the \(\beta\) band in PD4. In contrast, DeepConvNet assigned the highest proportion to the \(\gamma\) band in all four rats, with a higher \(\gamma\)-band proportion than EEGNet in each rat. These results indicate that the relative contributions of frequency bands to classification varied across both rats and model architectures. In particular, the prominence of the \(\theta\) band in PD3 for EEGNet shows that classification-relevant features were not restricted to high-frequency bands and could also include low-frequency information.

\begin{figure}[!htbp]
\centering
\includegraphics[width=.95\linewidth,trim={7.2476bp 388.5188bp 14.0009bp 10.9971bp},clip]{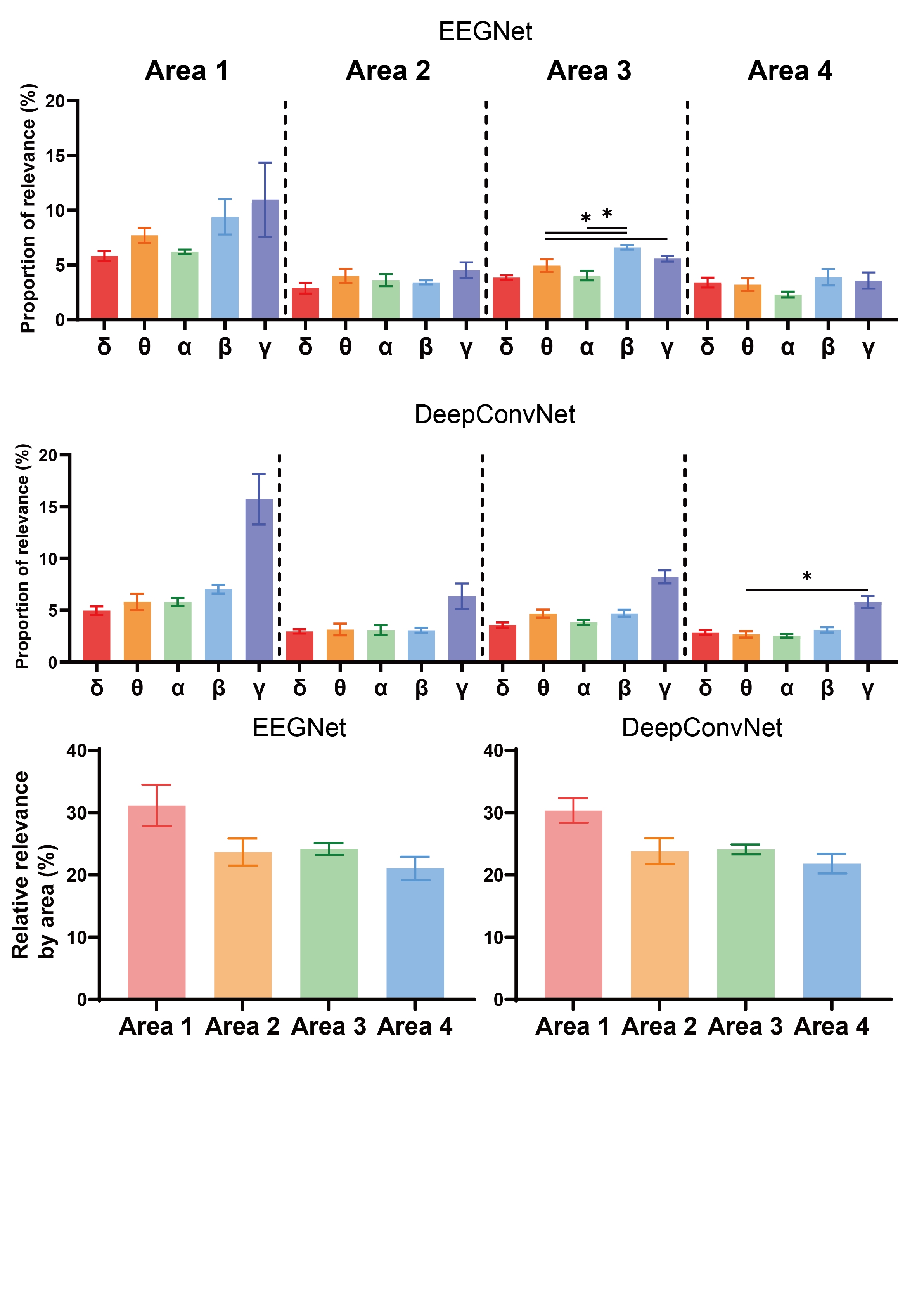}
\caption{Relative LRP relevance by cortical area and frequency band. Results are shown for EEGNet (top) and the reduced two-block DeepConvNet (bottom). Values represent the percentage of total absolute relevance assigned to each area--band combination in correctly classified post-stimulation test samples, without adjustment for the number of electrodes per area. Bars show the mean \(\pm\) SEM across four rats. Band effects within each area were evaluated using repeated-measures ANOVA with Greenhouse--Geisser correction and Bonferroni-adjusted paired comparisons. *p < 0.05.}
\label{fig:fig5}
\end{figure}

Figure 5 shows the percentage of total absolute relevance assigned to each area--band combination in correctly classified post-stimulation test samples. Values were calculated within each rat without adjustment for the number of recording channels per area and are presented as the mean \(\pm\) SEM across four rats. Band effects within each area were evaluated using repeated-measures ANOVA with Greenhouse--Geisser correction and Bonferroni-adjusted paired comparisons. For EEGNet, the relevance proportion in Area 3 was significantly higher in the \(\beta\) band than in the \(\delta\) and \(\alpha\) bands, and in the \(\gamma\) band than in the \(\delta\) band. For DeepConvNet, the relevance proportion in Area 4 was significantly higher in the \(\gamma\) band than in the \(\theta\) band. These comparisons indicate that the classifiers assigned greater relative importance to these high-frequency bands than to the corresponding lower-frequency bands within the respective areas when classifying post-stimulation samples.

\begin{figure}[!htbp]
\centering
\includegraphics[width=.78\linewidth,trim={5.0025bp 172.7620bp 17.2525bp 459.5199bp},clip]{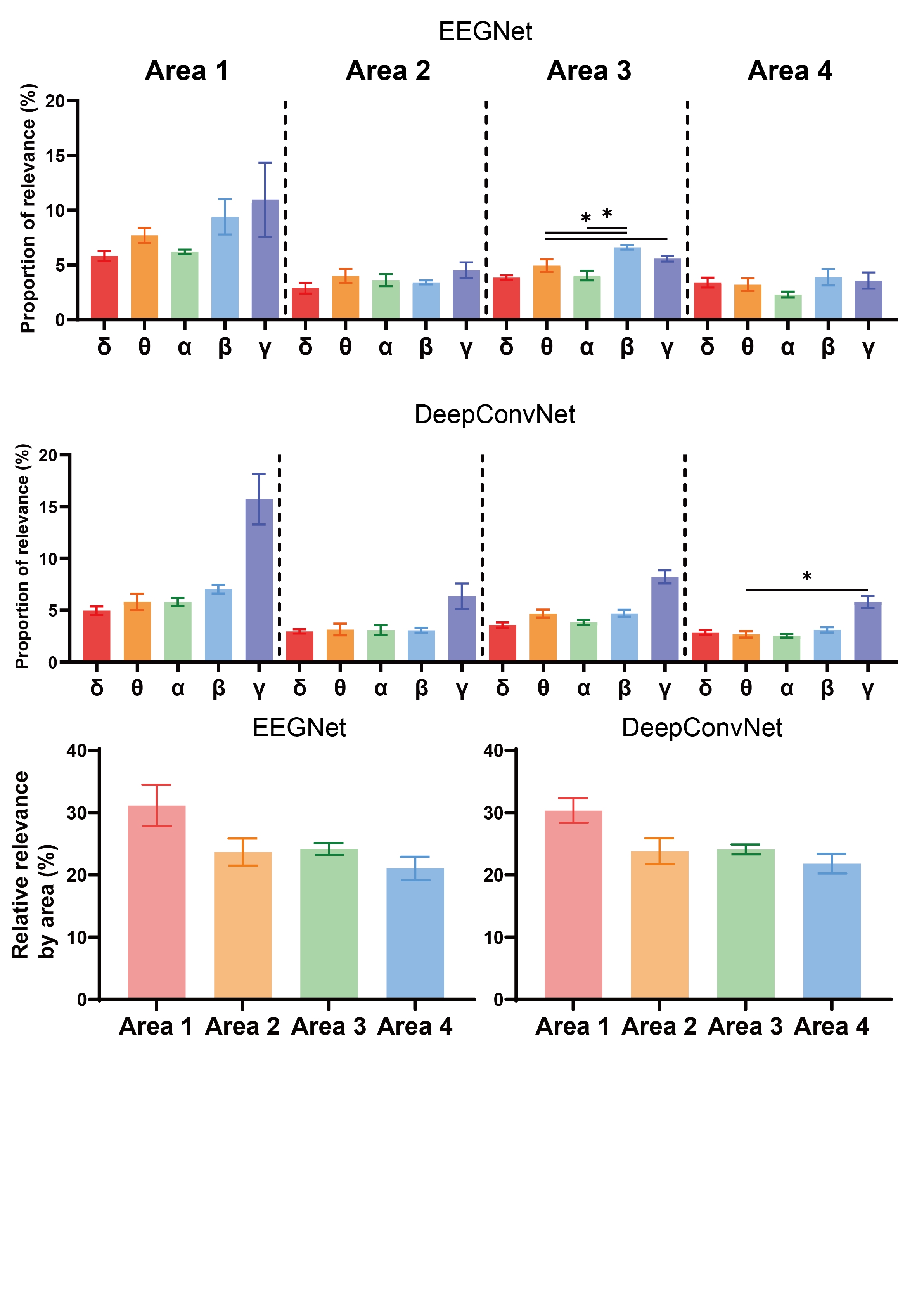}
\caption{Relative LRP relevance of cortical areas across frequency bands. Results are shown for EEGNet (left) and the reduced two-block DeepConvNet (right). For each rat and model, absolute relevance from correctly classified post-stimulation test samples was summed across all five frequency bands and recording channels within each cortical area. The summed relevance for each area was divided by the number of recording channels in that area to account for differences in channel count. These channel-count-adjusted values were then expressed as percentages of their sum across the four areas. Bars show the mean \(\pm\) SEM across four rats. Area effects were evaluated using repeated-measures ANOVA with Greenhouse--Geisser correction and Bonferroni-adjusted paired comparisons.}
\label{fig:fig6}
\end{figure}

Figure 6 shows the relative LRP relevance of the four cortical areas across all five frequency bands. For each rat and model, absolute relevance was summed across frequency bands and recording channels within each area and divided by the number of recording channels in that area. These values were then expressed as percentages of their sum across the four areas. At the area level, Area 1 showed the highest mean LRP attribution for both models, although the differences among areas were not significant. Both models highlighted the same cortical area despite their different network architectures, suggesting that features contributing to pre-/post-stimulation classification shared a common spatial component. These findings illustrate how XAI can identify regions containing classification-relevant features and reveal spatial similarities in the information used by different models.

\begin{table}[!htbp]
\centering\small
\caption{EEGNet classification accuracy after frequency-band masking without retraining. Values show the mean accuracy (\%) across five held-out test folds for each rat using 6-second inputs. Full denotes unmasked input. In the masked conditions, \(\beta\), \(\gamma\), or both bands in the standardized input were set to zero while model weights were fixed. All conditions were evaluated on the same test samples.}
\label{tab:table2}
\begin{tabular}{lllll}
\toprule[1pt]
Animal & Full (\%) & \(\beta\) masked (\%) & \(\gamma\) masked (\%) & \(\beta\)+\(\gamma\) masked (\%) \\
\noalign{\vskip\aboverulesep\begingroup\color[HTML]{808080}\hrule height 0.5pt\endgroup\vskip\belowrulesep}
PD1 & 98.390 & 74.650 & 75.790 & 46.490 \\
PD2 & 100.000 & 100.000 & 80.170 & 62.890 \\
PD3 & 100.000 & 100.000 & 35.570 & 33.790 \\
PD4 & 100.000 & 91.080 & 83.020 & 65.590 \\
\bottomrule[1pt]\end{tabular}
\end{table}

We evaluated changes in classification performance after masking the \(\beta\) and \(\gamma\) bands to further examine their contributions highlighted by LRP. The corresponding portions of the standardized input were set to zero while the trained EEGNet model weights were fixed, and all conditions were evaluated on the same test samples without retraining. As shown in Table 2, masking the \(\beta\) band decreased classification accuracy in PD1 and PD4 by 23.74 and 8.92 percentage points, respectively. These were also the two rats in which the \(\beta\) band had the highest relevance proportion in the EEGNet results (Figure 4). In PD2 and PD3, where \(\beta\)-band relevance was relatively lower than \(\gamma\)-band relevance, \(\beta\)-band masking did not decrease classification accuracy, whereas \(\gamma\)-band masking caused large decreases of 19.83 and 64.43 percentage points, respectively. Overall, masking the \(\gamma\) band caused a large decrease in performance in all rats, and masking both bands resulted in a larger decrease than masking the \(\gamma\) band alone. These findings provide support for the trained model's dependence on \(\beta\) and \(\gamma\) band information for classification, consistent with the rat-specific relevance patterns identified by LRP.

In summary, XAI allowed us to examine the frequency bands, regions, and recording channels that contributed to pre-/post-stimulation classification. The LRP results highlighted the contributions of the \(\beta\) and \(\gamma\) bands, and DeepLIFT yielded similar results (Supplementary Figure S3C). The involvement of high-frequency features is consistent with previous studies reporting pronounced changes in these bands following electrical stimulation in Parkinsonian rats \cite{ref4,ref24,ref25}. Both models also highlighted overlapping recording channels and a common cortical area. The masking results supported the trained model's dependence on \(\beta\) and \(\gamma\) band information for classification. Together, these analyses identified classification-relevant spatial and frequency features, providing complementary information about the features used to distinguish pre- and post-stimulation recordings alongside PSD-based analyses.

\subsection*{3.4. Effect of window size on the relevance scores}

\begin{figure}[!htbp]
\centering
\includegraphics[width=.85\linewidth,trim={27.7517bp 568.5308bp 31.0033bp 28.2506bp},clip]{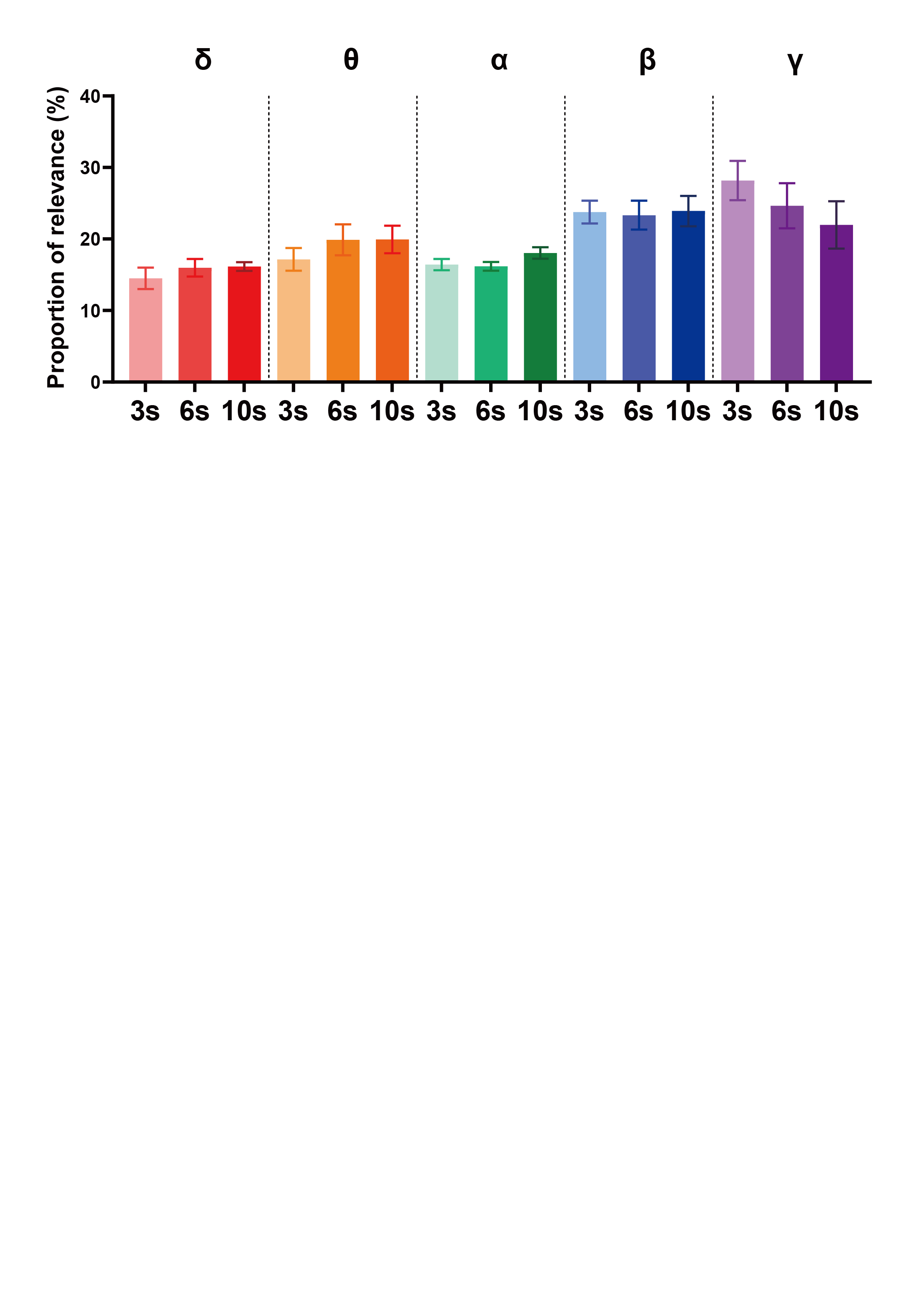}
\caption{Effect of input window length on relative LRP relevance across frequency bands. Band proportions are shown for 3-, 6-, and 10-second EEGNet inputs. For each rat and window length, absolute relevance from correctly classified post-stimulation test samples was summed within each band and divided by the total across all bands. The five band proportions sum to 100\%. Bars show the mean \(\pm\) SEM across four rats. Window-length effects within each band were evaluated using repeated-measures ANOVA with Greenhouse--Geisser correction and Bonferroni-adjusted paired comparisons.}
\label{fig:fig7}
\end{figure}

Changing the input window length changes both the temporal information available within each sample and the number of samples obtained from a recording. We therefore compared 3-, 6-, and 10-second EEGNet inputs to examine whether this choice affected the frequency-band relevance patterns identified by LRP. Figure 7 shows the percentage of total absolute relevance assigned to each frequency band. Relevance proportions were calculated from correctly classified post-stimulation test samples within each rat and are presented as the mean \(\pm\) SEM across four rats.

Window-length effects were evaluated within each frequency band using repeated-measures ANOVA with Greenhouse--Geisser correction. No statistically significant window-length effects were observed in any frequency band (Supplementary Table S6). This comparison extends the assessment of classification-relevant frequency features beyond the 6-second inputs used in the main analysis. Within the tested range, we found no statistical evidence that changing the input length altered the relative contributions of frequency bands, providing a sensitivity check for the interpretation of the XAI results.

\section*{4. Limitations}

A major limitation of this study is the small number of subjects. The invasiveness and complexity of surgical procedure and ECoG acquisition, high costs, and ethical concerns posed challenges in data collection from a large number of subjects. Consequently, we had to conduct our study using a subject-dependent approach, which is a key reason why our findings may not be generalizable to other subjects. Furthermore, assessing the importance of specific brain regions based on channel locations involves various factors, such as inter-subject difference (e.g., brain size) and subtle inconsistency during electrode placement. The limited number of subjects makes it difficult to ignore these variables, which may introduce bias into the analysis results. Additionally, our study lacked post-stimulation longitudinal behavioral assessments, which restricts the direct correlation between electrophysiological adaptations and functional motor recovery.

In addition, the ECoG dataset used in this study were collected in an earlier experiment designed to evaluate the effectiveness of electrical stimulation. In that experiment, behavioral assessments that included a control group confirmed improvements in motor function following electrical stimulation, but our ECoG dataset were not collected from the control group \cite{ref27}. Our ECoG dataset was collected over a long period without a control group. Therefore, differences in pre- and post-stimulation classification could be influenced by non-specific effects---such as electrode drift, tissue responses, and circadian factors---rather than electrical stimulation alone.

\section*{5. Conclusion}

This study combined graphene-based cortical stimulation with an explainable deep learning framework to characterize spatial and frequency features contributing to the classification of pre- and post-stimulation ECoG recordings. We applied EEGNet and DeepConvNet to ECoG signals acquired before and after electrical stimulation in Parkinsonian rats and used LRP and DeepLIFT to examine the features contributing to classification. The XAI analyses highlighted the contributions of \(\beta\)- and \(\gamma\)-band information and identified overlapping recording channels and a common cortical area emphasized by both models. The decreases in classification performance following \(\beta\)- and \(\gamma\)-band masking supported the trained EEGNet model's dependence on these frequency bands. By identifying classification-relevant spatial and frequency features, this approach provided complementary insights into the information used by the models to distinguish pre- and post-stimulation recordings alongside PSD-based analyses.

\section*{CRediT authorship contribution statement}

\textbf{Hanseul Choi:} Methodology, Software, Investigation, Writing - Original Draft. \textbf{Gaeun Kim:} Methodology, Formal analysis, Visualization, Writing - Original Draft. \textbf{Sunggu Yang:} Resources, Data Curation, Funding acquisition. \textbf{Eunha Baeg:} Data Curation, Writing - Review \& Editing. \textbf{Donggue Kim:} Formal analysis, Visualization, Writing - Review \& Editing. \textbf{Seongwon Jin:} Investigation. \textbf{Sangwon Byun:} Methodology, Validation, Writing - Review \& Editing. \textbf{Jibum Kim:} Conceptualization, Writing -- Review \& Editing, Supervision, Project administration, Funding acquisition. 

\section*{Declaration of competing interests}

The authors declare that they have no known competing financial interests or personal relationships that could have appeared to influence the work reported in this paper.

\section*{Data Availability}

The full dataset used in this study is not publicly available because of restrictions on data sharing imposed by the data owner. The source code used for preprocessing and XAI analysis, the trained model weights for each rat, and the test data from one evaluation fold corresponding to each model are publicly available on Zenodo at \url{https://zenodo.org/records/23717412}.

\section*{Funding}

This work was supported by the National Research Foundation of Korea (NRF) Grant funded by Korea Government through the Ministry of Science and ICT (MSIT) under Grant RS-2024-00341307. This work was also supported by the Institute of Information and Communications Technology Planning and Evaluation (IITP)-ICT Challenge and Advanced Network of HRD (ICAN) Grant funded by Korea Government (MSIT) under Grant IITP-2026-RS-2024-00437024.

\end{document}